# Using Isomorphic Problems to Learn Introductory Physics

Shih-Yin Lin and Chandralekha Singh
Department of Physics and Astronomy, University of Pittsburgh, Pittsburgh, PA 15260, USA

**Abstract**

In this study, we examine introductory physics students' ability to perform analogical reasoning between two isomorphic problems which employ the same underlying physics principles but have different surface features. 362 students from a calculus-based and an algebra-based introductory physics course were given a quiz in the recitation in which they had to first learn from a solved problem provided and take advantage of what they learned from it to solve another problem (which we call the quiz problem) which was isomorphic. Previous research suggests that the multiple-concept quiz problem is challenging for introductory students. Students in different recitation classes received different interventions in order to help them discern and exploit the underlying similarities of the isomorphic solved and quiz problems. We also conducted think-aloud interviews with four introductory students in order to understand in-depth the difficulties they had and explore strategies to provide better scaffolding. We found that most students were able to learn from the solved problem to some extent with the scaffolding provided and invoke the relevant principles in the quiz problem. However, they were not necessarily able to apply the principles correctly. Research suggests that more scaffolding is needed to help students in applying these principles appropriately. We outline a few possible strategies for future investigation.

**Introduction**

Learning physics is challenging. Physics is a subject in which diverse physical phenomena can be explained by just a few basic physics principles. Learning physics requires unpacking these principles and understanding their applicability in a variety of contexts that share deep features [1,2]. A major goal of most calculus-based and algebra-based introductory physics courses is to help students learn to recognize the applicability of a physics principle in diverse situations and discern the deep similarities between the problems that share the same underlying physics principles but have different surface features.

It is well known that two physics problems that look very similar to a physics expert because both involve the same physics principle don't necessary look similar to the beginning students [1]. Research has shown that when physics experts and novices are given several introductory physics problems and asked to categorize the problems based upon similarity of solution, experts tend to categorize them based upon the fundamental physics principles (e.g., conservation of mechanical energy, Newton's $2^{nd}$ Law, etc.) while novices tend to group them based upon the surface features such as pulley or inclined plane [1]. Similarly, when a group of introductory physics students and physics faculty were asked to rate the similarities between different pairs of problems, it was found that for problem pairs which only involve surface similarity but employ different principles, students were more likely to rate them as similar compared to the faculty members [3]. The different patterns that experts and novices discern in these problems reflects the difference between the ways in which the knowledge structure of experts and novices is structured and how they exploit it to solve problems. The fact that experts in physics have a well-organized knowledge hierarchy where the most fundamental physics principles are placed at the top, followed by layers of subsidiary knowledge and details facilitates their problem solving process, allowing them to approach the problems in a more effective and systematic way [1,2,4-13]. It also guides the experts to see the problems beyond the surface features, and makes the transfer of knowledge between different contexts easier.

There has been much research effort devoted to investigating and improving transfer of learning [14-26]. In these investigations, issues about transfer of knowledge from one context to another have been discussed from different perspectives [27-36]. The amount of knowledge a person has, the knowledge structure that the person constructs, and the context in which the knowledge is learned could all affect the person's ability to transfer knowledge acquired in one situation to another [27].

One way to help students learn physics is via analogical reasoning [1,2]. Students can be explicitly taught to make an analogy between a solved problem and a new problem, even if the surface features of the problems are different. In doing so, students may develop an important skill shared by experts: the ability to transfer from one context to another, based upon shared deep features. Here, we examine introductory physics students' ability to perform analogical problem solving. In this investigation, students were explicitly asked to point out the similarities between a solved problem and a quiz problem and then use the analogy to solve the quiz problem. In particular, students were asked in a recitation quiz to browse through and learn from a solved problem and then solve a quiz problem that has different surface features but the same underlying physics. Different types of scaffolding were provided in different intervention groups (recitation sections). The goal is to investigate what students are able to do with the analogy provided, and to understand if students could discern the similarities between the solved and the quiz problems, take advantage of them and transfer their learning from the solved problem to solve the quiz problem.

Our investigation also has overlap with prior investigations involving isomorphic problems since we focus on the effect of using isomorphic problem pair to help students learn introductory physics. In particular, students were explicitly asked to learn from a solved problem and then solve another problem which is isomorphic. According to Simon and Hayes [37], isomorphic problems are defined as problems that can be mapped to each other in a one-to-one relation in terms of their solutions and the moves in the problem solving trajectories. For example, the "tower of Hanoi problem" and the "cannibal and the missionary problem" are isomorphic to each other and have the same structure if they are reduced to the abstract mathematical form [37]. In this investigation, we call problems isomorphic if they can be solved using the same physics principles. The ballerina problem in which the ballerina's rotational speed changes when she pulls her arm closer to or farther away from her body is isomorphic to a neutron star problem in which the collapse due to gravity makes the neutron star spin faster. Both these problems require the conservation of angular momentum principle to solve them, but the contexts are very different.

Cognitive theory suggests that, depending on a person's expertise in the field, different contexts and representations may trigger the recall of a relevant principle more in one problem than another, and two problems which are isomorphic are not necessarily perceived as being at the same level of difficulty especially by a beginning learner [38,39]. Changing the context of the problem, making one problem in the isomorphic pair conceptual and the other quantitative, or introducing distracting features into one of the problems can to different extent raise the difficulty in discerning the similarity and make the transfer of learning between the two problems more challenging [40]. A previous study on transfer in which isomorphic problem pairs in introductory physics were given back to back to the students suggests that those who were given both the quantitative and conceptual problems in the isomorphic pairs were often able to perform better on the conceptual problem (which was typically more challenging for them) than the students who were given the conceptual problem alone [41]. For problem pairs that didn't involve a conceptual and a quantitative one but one problem provided a hint for the other, students typically were able to discern the similarity between the two problems and took advantage of what they learned from one problem to solve the other. However, for those problems in which the context triggered an alternative approach (which was not necessarily correct) to solve the problem (for example, in problems involving friction), the alternative view prevented the students from making a connection between the two problems. This study suggests that isomorphic problem pairs may be a useful tool to help students learn physics, but in some cases, more scaffolding may be needed [42].

As noted earlier, the study here could also be viewed from a broader perspective of learning and reasoning by analogy. Analogy is often useful in helping people understand an unfamiliar phenomenon. Theories suggest that analogy can make the mental processing of new information more efficient by modifying the existing knowledge schemata [43,44]. Similar to Piaget's idea of accommodation process, new schema can be created by transferring the existing cognitive structure from the source domain to the target domain in which analogy comes into play [43,44]. As pointed out in the literature [44], a good analogy not only creates an efficient connection between the new and existing information, but can also make the new information more concrete and easier to comprehend. Analogy can also be made by drawing a connection between different contexts involving similar reasoning strategies, e.g., in problems where the same physics principles are applicable, which is what we aim at here. The view of how analogy plays a role in the learning process which involves connecting the new material with the existing structure and modifying the existing cognitive structure to accommodate the new information is consonant with the view which describes learning as a construction process, emphasizing the importance of prior knowledge as a basis of learning. Studies have shown that using analogy could help improve students' learning and reasoning in many domains [43-47], and it has long been an effective strategy adopted by many instructors in the practical classrooms.

Another important thread of research related to the study discussed here is that of learning from examples. Examples can serve a goal similar to that served by analogy because they can be used to draw connection between different materials and make the unfamiliar familiar [43]. Presenting students with examples to demonstrate the meaning and application of a physics concept is a very common pedagogical tool in physics. Research on learning from worked-out examples [48-52] (such as those in a textbook) has shown that students who self-explain the underlying reasoning in the example extensively learn more than those who don't self-explain even if the self-explanations given by the students are sometimes fragmented or incorrect. It is suggested that the largest learning gain can be achieved if students are actively engaged in the process of sense making while learning from examples [48,50-52].

**Methodology**

In this study, students from a calculus-based and an algebra-based introductory physics course were given two isomorphic problems in the recitation quiz. The solution to one of the problems (which we call the "solved problem") was provided.

Students were explicitly asked to learn from the solution to the solved problem, point out the similarities between the two problems, explain whether they can use the solved problem to solve the other problem (which we call the "quiz problem"), and then they were asked to solve the quiz problem. The solution provided was presented in a detailed and systematic way. It started with a description of the problem with the knowns, unknowns, and target quantity listed, followed by a plan for solving the problem in which the reasons why each principle was applicable were explicated. After the plan was executed in the mathematical representation, the last part of the solution provided a check for the answer by examining the limiting cases. A full solution to the solved problems can be found in the supplementary materials.

In the quiz, the solved problem was about a girl riding on a rollercoaster car on a smooth track. The problem asked for the apparent weight of the girl when the car went over the top of a hump around which the track was part of a circle. Conservation of mechanical energy can be used to find the speed at the point of interest, followed by the application of Newton's $2^{nd}$ Law in the non-equilibrium situation with a centripetal acceleration to solve for the normal force, which is related to the target variable. This problem was isomorphic to the quiz problem, which was about a boy swinging on a tire swing created by a rope tied to a branch. Students were told that the rated maximum value of tension that the rope could hold was 2500 N. They were asked to evaluate whether the ride was safe by solving for the maximum tension in the rope during the ride, assuming the boy initially started at rest at a certain height. Again, the problem can be solved using the principles of conservation of mechanical energy and Newton's $2^{nd}$ Law as well as the concept of centripetal acceleration. The same problems have been used in another study, which examines the effect of students' self-diagnosing of their own solutions to quiz problems on subsequent problem solving and transfer [52,53]. In that study, students were asked in the quiz to solve the rollercoaster problem first, and then diagnose their own mistakes with different types of scaffolding provided to aid the self-diagnosis process. The swing problem was later given in the midterm exam. Although the solution to the swing problem can be mapped to that of the rollercoaster problem in an almost one-to-one fashion, many students didn't necessarily recall and transfer what they learned from the self-diagnosing task and didn't perform well on the swing problem [52,53]. It is possible that the time separation between the quiz and the midterm exam as well as the different contexts of the two problems made it difficult for students to discern the deep connection between the two problems. By explicitly placing the two problems in a pair, providing students with a detailed solution to one problem and asking them to point out the similarities between the two problems before solving the quiz problem, our goal in this study is to examine whether such explicit hints can help them make better connections between the two problems and help them solve the quiz problem by learning from the solved problem.

362 students from an algebra-based and a calculus-based introductory physics course were involved in this study (181 students in each, respectively). In each course, students were randomly divided into one comparison group and three intervention groups based on the different recitation classes. Students in the comparison group were given only the quiz problem in the recitation quiz. Similar to a traditional quiz, students in this comparison group were asked to solve the quiz problem on their own with no scaffolding support provided. The performance of this group of students could help us understand what students were able to do without being explicitly provided a solved isomorphic problem to learn from.

Students in the other three intervention groups, on the other hand, were given an opportunity to learn from the solved isomorphic problem during the quiz. Our previous research [54] indicates that simply providing students with a similar solved problem doesn't necessarily help them because students may simply follow the procedures in the solution without thinking carefully about the deep similarity of the problems. In order to help students process through the analogy more deeply and contemplate issues which they often have difficulty with, different kinds of scaffolding were provided in addition to the solved problem to the students in different intervention groups.

In particular, students in the intervention group 1 were asked to take the first few minutes in the quiz to learn from the solution to the solved problem. They were explicitly told at the beginning of the quiz that after 10 minutes, they had to turn in the solution, and then solve two problems in the quiz: one of them would be exactly the same as the one they just browsed over (the rollercoaster problem), and the other one would be similar (the swing problem). In order to help students discern the connection between the two problems, students were also explicitly asked to identify the similarities between the two problems and explain whether they could use the similarities to solve the quiz problem before actually solving it. We hypothesized that since they had to solve the same problem whose solution they browsed over and another isomorphic problem in the quiz, students would try hard to get the most out of the solution in the allocated learning period. In order to apply what they learned from the solution to solve exactly the same problem on their own as well as an isomorphic problem, they had to not only focus on what principles are useful, but also understand why and how each principle is applicable in different circumstances. We hypothesized that an advantage could be achieved over the comparison group if students in the intervention group 1 went through a deep reasoning while browsing over the solved problem. Students' performance on both problems was later analyzed and compared with the comparison group.

The scaffolding in the 2nd intervention group was designed based on a different framework. Students in this group were first asked to solve the quiz problem on their own. After a designated period of time, they turned in their solution, and were given the isomorphic solved problem to learn from. Then, with the solved problem and its solution in their possession, they were asked to redo the quiz problem a second time after pointing out the similarities between the two problems and explicitly asked to discuss the implication of these similarities in constructing their solution to the quiz problem. We hypothesized that postponing the browsing over the solved isomorphic problem until the students have actually tried to solve the quiz problem on their own could be beneficial to them because in this way, students would have already searched through their knowledge base of physics and attempted to organize the information given in the quiz problem. We hypothesized that having tried the quiz problem on their own may make the browsing over the solved problem for relevant information more structured and productive before students attempted the quiz problem a second time. Students had the opportunity to display what they learned from the solved isomorphic problem when they solved the quiz problem a second time. The fact that the solution we provided had made explicit the consideration for using the principles but was not directly the solution to the quiz problem was inspired by Schwartz, Bransford and Sears' theory of transfer [35], which states that two components, efficiency and innovation, are both important in the learning process.

Unlike the students in the intervention groups 1 and 2 who had to figure out the similarities between the two problems themselves, students in the 3rd intervention group were given both the quiz problem and the solved problem at the same time and were explicitly told that "Similar to the solved problem, the quiz problem can be solved using conservation of energy and Newton's 2nd Law (with centripetal acceleration)". We hypothesized that deliberately pointing out the principles that are useful in solving both problems may guide students to focus more on the deep physics instead of the surface features while browsing over the solved problem. In addition to the instruction which asked them to first learn from the solved problem and then exploit the similarity to solve the quiz problem, students in this group also received extra hints to help them deal with the common difficulties in solving this problem found in previous research [54-56].

Research suggests that introductory physics students have great difficulty dealing with the non-equilibrium situation and they usually think of a non-equilibrium situation which involves the centripetal acceleration as an equilibrium situation by treating the centripetal force as an additional force [56]. In the swing problem, the correct use of the centripetal acceleration and Newton's 2$^{nd}$ Law should yield $T - mg = \frac{mv^2}{r}$. However, students who treat it as an equilibrium problem and believe that "the centripetal force is an additional force" obtain an answer of the type $T - mg + \frac{mv^2}{r} = 0 \Rightarrow T - mg = -\frac{mv^2}{r}$, which has a wrong sign. To help students with these issues, we presented to students in the intervention group 3 a dialogue between two people discussing whether the centripetal force is an additional force or whether it is simply a name given to the net force in a circular motion. (See the supplementary material.) Students were asked to explain which person they agreed with and why before solving the quiz problem. To assist students in correctly analyzing the dialogue, a practical situation similar to the rollercoaster cart, which went over the top of a circle was discussed. Free-body diagrams as well as mathematical equations were presented with the dialogue. We hypothesized that if students did not know how to assess which person is correct in the dialogue, they could always go back to the solution of the rollercoaster problem provided and figure out the correct answer by comparing either the free-body diagrams or the mathematical equations. We hypothesized that after students contemplated the issues discussed in the dialogue and acquired a better understanding of the centripetal acceleration and centripetal force, they may perform better on the quiz problem.

Students' performance on the quiz was graded by two researchers using a rubric. Summary of the "physics part of the rubric highlights" for the quiz problem is shown in Table 1. The rubric for the solved problem is not listed here because the solutions to the two problems can be mapped directly to each other and the rubrics are almost identical except for the problem specific details involving the application of physics principles. As shown in Table 1, the rubric had a full score of 10 points, divided into two parts based upon the two principles involved. Three points were devoted to using the principle of conservation of mechanical energy (CME) to find the speed at the point where Newton's 2$^{nd}$ Law was applied; seven points were devoted to identifying the centripetal acceleration, recognizing all relevant forces and applying Newton's 2$^{nd}$ law correctly to obtain the final answer. Students' common mistakes and the corresponding points taken off are also listed. In the case of intervention 3, which included an additional dialogue problem, the same rubric was used for grading their answer for finding the tension. If the students didn't answer the dialogue problem correctly, an additional 2 points were taken off from the score they received for solving for the tension force if it didn't result in a negative score. The minimum score was zero. An inter-rater reliability of more than 80 percent was achieved when two researchers scored independently a sample of 20 students. When five researchers scored independently a sample of five students, the inter-rater reliability was more than 95 percents.

Table 1 Summary of the rubric for the quiz problem. The rubric for the solved problem is almost identical.

| Description | Correct answer | Common mistakes | | Points taken off |
|---|---|---|---|---|
| Invoking and applying the principle of conservation of mechanical energy to find the speed (3 points) | $mg\Delta h = \frac{1}{2}mv^2$ | use 1-D kinematics equations to find $v$ | | 2 |
| | | wrong $\Delta h$ | | 1 |
| Identifying the centripetal acceleration and using Newton's 2nd Law to find the tension (7 points) | $a = a_c = \frac{v^2}{r}$ $T - mg = ma_c = m\frac{v^2}{r}$ $\Rightarrow T = mg + m\frac{v^2}{r}$ | $a = 0, \quad T = mg$ | | 5 |
| | | $a \neq 0$ but wrong formula for $a$ | $2500 = ma$ | 3 |
| | | | $a = \frac{mv^2}{r}$ | 1 |
| | | $T = m\frac{v^2}{r}$ | | 3 |
| | | $T = mg - m\frac{v^2}{r}$ | Using $\sum F = 0$ (centripetal force as an additional force) | 2 |
| | | | Using $\sum F = T - mg$ $= ma = -\frac{mv^2}{r}$ (didn't pay attention to the direction of $a$) | 1 |

Students' performance in different intervention groups was later compared to each other. In order to examine the effects of interventions on students with different expertise and to evaluate whether the interventions were more successful in helping students at a particular level of expertise, we further classified the students in each course as top, middle, and bottom based on their scores on the final exam. Students in the whole course (no distinction between different recitation classes) were first ranked by their scores on the final exam. About 1/3 of the students were assigned to the top, middle, and bottom groups, respectively. The overall performance of each intervention group is represented by an unweighted mean of students' performance from the three different levels of expertise. To compare how similar the students in different intervention groups were, their performance on the Force Concept Inventory (FCI) [57] administered at the beginning of the semester was investigated. There was no statistically significant difference between the different intervention groups in terms of the FCI score. Moreover, in order to take into account the possible difference which may develop as the semester progresses between different recitation classes, the effects of different interventions on the quiz were compared based on the unweighted means described earlier.

In addition to the comparison between the different intervention groups, we also compared the students' performance in these algebra-based and calculus-based introductory physics courses with the performance of a group of first-year physics graduate students who were asked to solve the tire swing problem on their own without any solved problem provided. The performance of the graduate students can serve as a benchmark for how well the undergraduate students can perform. Moreover, we also conducted think-aloud interviews with four introductory physics students (who were selected from other introductory physics classes) to get an in-depth account of their difficulties with the scaffolding provided and examine additional ways to help them. The details of the interviews will be discussed later.

## Results and Discussion

### Quantitative Data from the two introductory physics courses

We found that the similarities between the solved and quiz problems that the students described in the first part of their quiz solution had no correlation with their ability to actually solve the quiz problem. Many students described the similarities based on the details of the problems (e.g., the initial speeds in both problems are zero, both problem are asking for a force, etc.) whether or not they could solve the problem correctly. For example, one student who correctly solved the quiz problem described the following three similarities: "*<1> going around a circle with m (30kg) and radius (15m) <2> need to solve for velocity at a point <3> start from rest*". However, the student did not mention the deep similarities regarding the physics principles involved. In particular, without looking at his actual solution to the quiz problem, it is not possible to tell whether this student would be able to solve the quiz problem correctly. On the other hand, the fact that some students described the similarities in terms of the physics principles involved didn't necessarily mean that they knew how to apply the principles correctly, and sometimes they didn't even make use of the principles they mentioned as similar (for the solved and quiz problems) when solving the quiz problem. For example, one student described the similarities as follows: "*The initial velocity of both is 0 m/s. The theory of conservation of energy is used in both. The tension is going to be $T = mg + \frac{mv_B^2}{R_B}$ instead of $N_B = mg - \frac{mv_B^2}{R_B}$. This is because in problem 1, the cart is moving up whereas in problem 2, the swing is moving*

*downwards in the arc, so the forces are acting as one combined force.*" Although these statements about the similarity seem to indicate that this student was capable of solving the quiz problem correctly, examination of his actual work shows that he didn't make use of the principle of conservation of energy at all in his actual attempt to solve the quiz problem. Instead, he tried to find the speed at the bottom of the ride by connecting the centripetal acceleration to the acceleration due to gravity and set $a_c = \frac{v^2}{r} = g = 10 \text{ m/s}$. Because of such inconsistencies, in the following discussion, we will only focus on students' *solutions* to the quiz problem (and not focus on their response to the question asking for the similarities between the two problems).

Table 2 and Table 3 present students' average scores on the tire swing problem (the quiz problem) in the calculus-based and algebra-based courses respectively. For the intervention group 2, students' performance when they solved the problem the 2$^{nd}$ time is presented. Due to the instructor's time constraint in the recitation classes, the allotted time for students in intervention group 2 to try the quiz problem on their own before learning from the solved problem was slightly less than the time given to those in the comparison group. Therefore, instead of examining how intervention 2 students' pre-scaffolding performance compares to that of the comparison group, we only focus on the performance of students in intervention group 2 after the scaffolding support was provided. Moreover, as noted earlier, the initial FCI scores were comparable for the comparison group and all intervention groups.

The p-values presented in Table 4 show that all three intervention (Intv) groups in the algebra-based course and the intervention group 2 in the calculus-based course significantly outperformed the comparison group, indicating that these students, to a moderate extent, could reason about the similarities between the two problems and take advantage of the solved problem provided to solve the quiz problem. On the other hand, while the score of the intervention group 1 in the calculus-based course was higher than the comparison group in the same course, the difference is not statistically significant. The performance of intervention 3 students in the same course was comparable to that of the comparison group. It is possible that many students in these groups failed to process the analogy between the solutions to the solved and quiz problems deeply the way we had hypothesized. We'll describe the possible reasons for the difficulty in analogical reasoning in the later paragraphs.

The algebra-based students benefited more from the interventions overall in the sense that students in all three intervention groups in general performed significantly better than the comparison group students. However, comparison of the absolute scores of students in the same intervention group from the two courses indicates that the calculus-based students on average scored higher than the algebra-based students whether or not the scaffolding was provided. We note that how well a student performed may depend not only on the scaffolding provided, but also on their initial knowledge relevant for the problem. An improvement would easily be seen if the students who initially had no clue about how the solution should be constructed were able to invoke an appropriate concept or principle by learning from the isomorphic problem provided. The fact that 26% of students in the algebra-based comparison group received a score of zero because they incorrectly connected the tension force directly to the energy (for example, with the equation $T = mgh$) suggests that there was plenty of room for improvement in invoking the principles correctly. A noticeable progress would be made if the students were able to recognize the similarity between the solved and quiz problems and identify correctly the principles to be used. However, in order to apply the physics principles successfully, more understanding and mathematical competence is required and students must also be able to understand the nuances between the solved and quiz problems.

Table 2 Students' average scores out of 10 on the tire swing problem (the quiz problem) in the calculus-based course. The number of students in each case is shown in parentheses. The performance of the whole group taken together is represented by an unweighted mean of students' average scores from the top, middle and bottom categories.

|         | Comparison (38) | Intv 1 (35) | Intv 2 (34) | Intv 3 (74) |
|---------|-----------------|-------------|-------------|-------------|
| Top     | 8.6 (14)        | 9.3 (15)    | 9.2 (13)    | 7.6 (19)    |
| Middle  | 7.6 (10)        | 8.7 (9)     | 9.4 (12)    | 7.5 (35)    |
| Bottom  | 4.2 (14)        | 4.6 (11)    | 8.7 (9)     | 5.1 (20)    |
| Average | 6.8             | 7.5         | 9.1         | 6.7         |

Table 3. Students' average scores out of 10 on the tire swing problem in the algebra-based course. The number of students in each case is shown in parentheses. The performance of the whole group taken together is represented by an unweighted mean of students' average scores from the top, middle and bottom categories.

|        | Comparison (54) | Intv 1 (46) | Intv 2 (33) | Intv 3 (48) |
|--------|-----------------|-------------|-------------|-------------|
| Top    | 6.0 (19)        | 8.0 (10)    | 6.8 (12)    | 7.2 (27)    |
| Middle | 2.7 (15)        | 7.3 (20)    | 6.7 (10)    | 3.5 (11)    |
| Bottom | 2.0 (20)        | 6.6 (16)    | 4.8 (11)    | 6.2 (10)    |
| Average| 3.5             | 7.3         | 6.1         | 5.6         |

Table 4. The p values (from ANOVA) for the comparison of students' performance between different groups in the calculus-based and algebra-based courses. The "c" stands for the comparison group.

|          | c vs. 1 | c vs. 2 | c vs. 3 | 1 vs. 2 | 1 vs. 3 | 2 vs. 3 |
|----------|---------|---------|---------|---------|---------|---------|
| Calculus | 0.200   | 0.000   | 0.829   | 0.091   | 0.417   | 0.000   |
| Algebra  | 0.000   | 0.000   | 0.001   | 0.417   | 0.371   | 1.000   |

For comparison, Table 5 lists the different answers graduate students provided to the tire swing quiz problem on which they achieved an average score of 8.4 out of 10. Out of the 26 graduate students, 21 students successfully figured out the correct answer. The most common mistakes the graduate students made were ignoring the fact that there was an acceleration involved and treating the problem as an equilibrium problem. A similar result has been reported [58] when the same problem was given to a group of physics professors.

In our study with the introductory physics students here, not recognizing the existence of the acceleration was one of the common mistakes, but this was not the only difficulty introductory students had. Without the interventions, some students (especially in the algebra-based course) simply had no clue about how to solve the problem and they tried to associate the tension force with some quantity that didn't even have the same dimension. Some students realized that they should apply Newton's 2$^{nd}$ law in the non-equilibrium situation but they didn't know how to find the acceleration. Even if some of them knew the expression for magnitude of the acceleration as $a_c = \dfrac{v^2}{r}$, they didn't necessarily know how to find the speed of the object. These difficulties, as well as the mistake of neglecting the gravitational force term in the solution, were reduced after the students were provided with the solved problem. With the scaffolding, more students were able to identify the existence of both the gravitational force and the centripetal acceleration, and most students could apply the principle of CME to find the speed correctly.

Table 5 Graduate students' answers to the tire swing problem.

| Answers | Number of people |
|---|---|
| $T = mg + \dfrac{mv^2}{r}$ (correct) | 21 |
| $T = mg$ | 4 |
| $T\cos\theta = mg$, $T\sin\theta = \dfrac{mv^2}{r}$ | 1 |

Examining intervention 2 students' performance shows that students did improve significantly by learning from the isomorphic solved problem provided after struggling with the quiz problem first. In particular, this intervention worked very well for the calculus-based students. With the solved problem in their possession to learn from, the calculus-based students achieved an average score of 9.1 (out of 10) the second time they solved the quiz problem, which was a higher score than the benchmark (8.4) set by the graduate students. Even the bottom students in this group earned an average score of 8.7 out of 10. Table 6 provides insight on how the pre and post performance of this group of students evolved by binning the students into different categories based on their solutions. A comparison of the number of students who had difficulty figuring out the acceleration and the speed correctly before and after the scaffolding was provided is shown in

Table 7. These tables suggest that most calculus-based students were able to correctly invoke the necessary knowledge which they lacked initially. They corrected at least part of their mistakes after browsing over the solution, and a significant improvement in the scores was found.

Table 6 Different answers calculus-based intervention 2 students provided for the tire swing problem before and after the scaffolding was provided. The corresponding number of students in each case is listed. The correct answer is indicated by the shaded background.

| | Before | After |
|---|---|---|
| $T = mg + \dfrac{mv^2}{r}$ | 13 (38.2%) | 26 (76.5%) |
| $T = mg - \dfrac{mv^2}{r}$ or $T = -mg + \dfrac{mv^2}{r}$ | 3 (8.8%) | 4 (11.8%) |
| $T = \dfrac{mv^2}{r}$ | 3 (8.8%) | 3 (8.8%) |
| $T - mg = ma$ but didn't know how to find $a$ | 3 (8.8%) | 0 |
| $T = mg$ | 4 (11.8%) | 0 |
| Other (e.g., $T = mv_f$) | 8 (23.5%) | 1 (2.9%) (This person thought $T_{max}$ occured when $\theta = 45^o$ and said $T_{max} \cos\theta - mg = \dfrac{mv^2}{r}$) |

Table 7 Comparison of the number of students who had difficulty figuring out the acceleration and the speed correctly before and after the scaffolding was provided in the calculus-based intervention group 2.

| | Before | After |
|---|---|---|
| Mentioned $a$ but had no idea how to find $a$ or used incorrect method to find $a$ (e.g., used $ma = 2500$ N to find $a$) | 4 | 0 |
| Used incorrect method to find $v$ (e.g., $v = 0$, $v = 9.8$ m/s, $v^2/r = g$, using 1-D kinematics equations) | 12 | 1 (used 1-D kinematics equations) |

Table 8 presents intervention 1 students' performance on the rollercoaster problem right after learning from and returning its solution to the instructor. It shows that many students in both the calculus-based and algebra-based courses were capable of reproducing the solved problem immediately. The average scores on the solved problem reproduced from students with different levels of expertise were 8.5 (calculus) and 9.0 (algebra); even the scores of the "bottom" students in both courses were high. The fact that students were immediately able to reproduce the problem they browsed over, however, doesn't necessarily mean that they could transfer their learning to a new isomorphic problem. An average drop of 1.0 and 1.7 points were found for the calculus-based and algebra-based students for the transfer problem. In fact, the "bottom" calculus-based students' average score on the quiz problem dropped to 4.6. One possible reason for this low score is that this group of students might not have as strong a motivation to perform well as the algebra-based students, and they didn't process through the solutions provided as deeply as we had hypothesized. The fact that these "bottom" students in the calculus-based course didn't perform well on the quiz problem as compared to other students who received the same intervention could be a possible reason for why on average the score of the intervention 1 students in the calculus-based course was not significantly better than the comparison group students.

Table 8  Average scores out of 10 on the roller coaster problem (solved problem) and the tire swing problem (quiz problem) for intervention 1 in the algebra-based and calculus-based courses. The performance of the whole group is represented by an unweighted mean of students' average scores from the top, middle and bottom categories.

|  | Solved Problem | | Quiz Problem | |
| --- | --- | --- | --- | --- |
|  | Calculus | Algebra | Calculus | Algebra |
| Top | 9.0 | 9.6 | 9.3 | 8.0 |
| Middle | 8.7 | 9.0 | 8.7 | 7.3 |
| Bottom | 7.9 | 8.5 | 4.6 | 6.6 |
| Average | 8.5 | 9.0 | 7.5 | 7.3 |

Although the solved problem provided was useful in helping students construct an appropriate solution plan for the quiz problem by invoking the relevant principles and correcting the terms they might have missed before browsing over the solved problem, students weren't necessarily able to apply the principles correctly when a change in the details of application was required in order to solve the transfer problem in the new situation. One common incorrect answer intervention 1 and 2 students provided (after learning from the solved problem) for the swing problem was $T = mg - \frac{mv^2}{r}$ (or sometimes $T = -mg + \frac{mv^2}{r}$ if the students noticed that the former answer would result in a negative value) instead of the correct answer of $T = mg + \frac{mv^2}{r}$. One possible reason for this mistake may be that the vector nature involved in Newton's 2$^{nd}$ law was challenging for the students. To apply the principle correctly, students need to realize that when applying Newton's 2$^{nd}$ law, not only do they have to take into account the direction of the forces, they also must remember the fact that the acceleration is also a vector in which a positive or negative sign based on the direction should be considered and assigned accordingly. If the students didn't realize that the centripetal accelerations were pointing in the opposite directions in these two problems (because in one problem the object was at the top and in the other, it was at the bottom) and they simply copied down the equations from the solved problem, they were likely to make the mistake.

Another possible reason for why students came up with a wrong sign for the centripetal acceleration term may be that they interpreted the quantity $\frac{mv^2}{r}$ as an additional force acting on the object undergoing a circular motion and they treated the situation as an equilibrium problem in which all the forces should sum up to zero. Intervention 3 students' answers to the additional dialogue question show that 30% and 35% of the calculus-based and algebra-based students, respectively, agreed with the first person who argued that "If an object is undergoing a circular motion, then there's an extra centripetal force acting on it" and that "If an object is traveling on a track of a vertical circle, using Newton's 2$^{nd}$ law in equilibrium situation, at the top we have $\sum F = 0 \Rightarrow N = mg + \frac{mv^2}{r}$." (See the supplementary material.) However, examination of students' work indicates that students were not always consistent between the answers they chose for the dialogue question and the actual solution they provided for the tire swing problem. The answers "agreeing with person 1" and "$T = mg - \frac{mv^2}{r}$" should be correlated if the students were consistent. Another consistent answer pair would be "agreeing with person 2" and "$T = mg + \frac{mv^2}{r}$".

Table 9 lists the intervention 3 students' answers to the dialogue question and the tire swing problem; the consistent answer pairs are indicated by the shaded background. The table suggests that a large fraction of the students were not consistent in their answers in both the algebra-based and calculus-based courses. It appears that some students didn't understand the key points in the two arguments and incorrectly agreed with one person based on some subsidiary factor. A student who correctly proceduralized Newton's 2$^{nd}$ Law in the non-equilibrium situation and came up with a correct answer agreed with person 1 "because centripetal force points into the center of the circle" (despite the fact that person 2 had a similar statement of "centripetal acceleration's direction is pointing from the object to the center of the circle." ). It is also likely that some students chose the inconsistent answer pairs because they expected the dialogue question to be directly applicable to the tire swing problem to be solved and they didn't recognize that these two cases involved different situations and different application details (since in the dialogue, the object was at the top but in the swing problem it was at the bottom). They either directly copied the final answer from the person they agreed with in the dialogue as their answer to the tire swing problem without thinking through it in the new situation, or they first solved for the tension in the tire swing problem and argued that whichever person had the same equation as theirs (if the normal force in the dialogue situation was substituted by the tension force in the quiz problem) would be the one they agreed with. In either of these cases, students lost 1~2 points because the person they agreed with reflected gaps in their knowledge structure (listed in the $2^{nd}$ last item of the rubric) or because the equation they used from the dialogue had a wrong sign (which would not have happened if they used correct concepts and derived the equation in the new situation themselves). The fact that some students lost additional points for the answer they gave to the dialogue question is one of the reasons why students in the intervention group 3 didn't perform as well as students in the other intervention groups. Another reason may be that providing students with more hints, e.g., by directly telling them the principles involved, may have reduced the amount of cognitive engagement and students may not be as actively involved in the reasoning in intervention 3.

Comparing the performance of different intervention groups, we found that all three intervention groups were significantly better than the comparison group in the algebra-based course and there was no significant difference between any of the intervention groups. In the calculus-based course, intervention 2 was the only group which statistically significantly outperformed the comparison group. It was also statistically significantly better than intervention group 3. As described earlier, the interventions would be useful if the scaffolding supports provided matched well with students' abilities and if the students were actively engaged in the thinking process as hypothesized during the design of each intervention. We found that to begin with, many algebra-based students had no clue about how to construct the quiz problem. Providing them with the solved problem (regardless of the different interventions) did help them invoke the relevant principles and an improvement was observed. As for the calculus-based students, whose initial performance was better, the intervention which let them struggle first before any scaffolding was provided benefited them the most. It is likely that this intervention was the one which made students think through the analogy between the solved and quiz problems with the greatest depth because the struggling experience can make students aware of their initial knowledge explicitly. Comparing what they learned from the solved problem with what they had initially thought, they had a good probability of detecting any discrepancy between them and were more likely to be forced to think about how to modify their initial knowledge and incorporate the new information to their existing knowledge structure in a coherent way. It is possible that students in the other two intervention groups were not forced to go through the analogy in great depth and some of them didn't think through the analogy between the solutions the way we had hypothesized. We'll describe the students' responses to interventions 1 and 3 further in the interview section.

Table 9 Intervention 3 students' answers to the dialogue question and the tire swing problem and the corresponding number of students in each case. The consistent answer pairs are indicated by the shaded backgrounds. In the calculus-based course, there were only 73 students in total because one student who answered that he "agreed with either student 1 or 2" was not included in this table.

|  | Calculus | | Algebra | |
| --- | --- | --- | --- | --- |
|  | Person 1 | Person 2 | Person 1 | Person 2 |
| $T = mg + \dfrac{mv^2}{r}$ | 17 | 19 | 7 | 10 |
| $T = mg - \dfrac{mv^2}{r}$ or $T = -mg + \dfrac{mv^2}{r}$ | 2 | 19 | 4 | 12 |
| $T = \dfrac{mv^2}{r}$ | 0 | 3 | 0 | 1 |
| $T = mg$ | 0 | 1 | 1 | 3 |
| Other | 3 | 9 | 5 | 5 |

**Interview- General Description**

In addition to the students from the previously discussed calculus-based and algebra-based courses who took the quiz, four students from several other introductory physics classes were recruited for one-on-one interviews to get an in-depth account of their reasoning while they solved the problems. Two of the four students we interviewed were enrolled in an algebra-based introductory mechanics course at the time of the interview; the other two were enrolled in two different calculus-based mechanics courses. The interviews were conducted after all the relevant topics had been covered in the lectures. All four students recruited had a midterm score which fell in the middle of their own introductory physics course, ranging from +3 to -9 points above or below the class averages (which fell between 70% and 76% for different sections of the courses). The audio-recorded interviews were typically 0.5-1 hour long.

During the interviews, students were asked to learn from the solved problem provided and solve the isomorphic quiz problem given. Different students received different kinds of interventions in the interviews, which are listed in Table 10. Most of the interventions were the same as the previous interventions used in the quantitative data discussed in the earlier section. One of them (what student A received) was new in the sense that a slight modification was made to the interventions used earlier. Instead of letting student A read the rollercoaster problem on his own and reproduce the rollercoaster problem again, the researcher outlined the solution to the solved problem to the student. After the student understood how to solve the rollercoaster problem, the researcher then asked him to solve the tire swing problem (quiz problem).

Table 10 The interventions students received in the interview.

|  | Quiz 1 |
| --- | --- |
| Student A | Modified Intv 1 |
| Student B | Intervention 1 |
| Student C | Intervention 3 |
| Student D | Intervention 3 |
| * Modified Intv 1 for quiz 1: (1) The researcher first discussed with the student how to solve the rollercoaster problem using Newton's $2^{nd}$ law and the reason why there is a minus sign in the centripetal acceleration term (2) The student looked at the solution to the solved problem for a short period of time (3) The student attempted to solve the quiz problem. | |

The interviews were conducted using a think-aloud protocol to follow and record the students' thinking processes. Students were asked to perform the task (whether they were reading the solved problem or trying to solve the quiz problem) while thinking aloud; they were not disturbed during the task. After the students completed the quiz, the researcher would first ask clarification questions in order to understand what they did not make explicit earlier and what their difficulties were. Based on this understanding, the researcher then provided some guidance (sometimes including the physics knowledge required) to the students in order to help them solve the quiz problem correctly if they had not done so. After helping students learn how to solve the quiz problem correctly, the researcher invited them to reflect on the learning process they just went through (for example, by asking explicitly what was the thing that helped them figure out how to solve the problem) and provide some suggestion from the student's own perspective on how to improve students' performance on the problem. The goal of the students' reflection was to help us identify the possible helpful scaffoldings not only based upon what the researchers observed but also based upon students' reflection of their own learning.

**Interview Results**

We found that many of the student difficulties observed in the quantitative data were observed in the interviews as well. In the following section, we will discuss some findings from the interviews which provided more in-depth understanding of students' thinking processes. Some check points that are likely to provide guidance to the students in successfully solving the tire swing problem will be summarized at the end.

First of all, we found that some students didn't take advantage of the solved problem to think through the analogy in a great depth as we had hoped. When designing intervention 1, we hoped that students will not only learn from the solved example regarding what principles should be invoked and why but also how the principles should be applied. We also hoped that requesting students to reproduce the solved problem could give them an opportunity to practice applying the principles before applying it to the quiz problem. When student B, who was given intervention 1, was instructed to solve the rollercoaster problem he just browsed over, he tried to reproduce the solved problem by simply recalling the equation he had just read. He didn't start from the fundamental principles to derive the equation, but rather simply wrote down the equations he remembered for the speed at the point of interest and the final targeted variable (which were both incorrect). His answers for the solved and quiz problems, which are displayed in Fig 1 and Fig 2, indicate that he superficially mapped the two problems together without carefully examining the differences. As this interview suggests, if the students didn't carefully think through the problems as we had hoped, it's less likely that they would benefit significantly from the interventions.

Fig 1 Student B's answer to the solved problem.

$M = 120$ kg   $h_A = 15$m   $h_B = 5$m   $R = 30$m
$m = 55$ kg

$$v_B = \sqrt{2gh_A + 2gh_B}$$
$$v_B = \sqrt{2(10)(15m) + 2(10 \text{ m/s}^2)(5m)}$$
$$v_B = 20 \text{ m/s}$$
$$N_B = mg + \frac{(M+m)v_B^2}{R^2}$$
$$N_B = 55kg\left(10\frac{m}{s^2}\right) + \frac{175kg\,(20\frac{m}{s})^2}{30m\,^2}$$
$$\boxed{N_B = 96.11 \text{N}}$$

Fig 2 Student B's answer to the quiz problem.

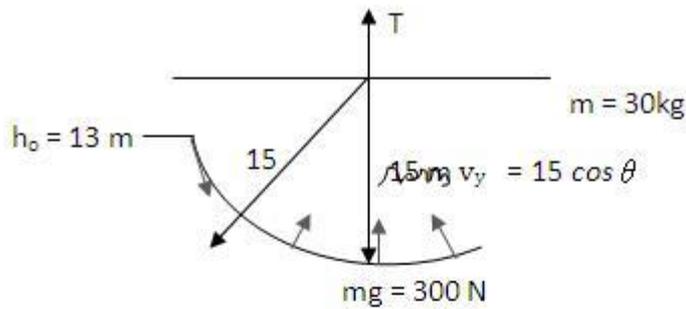

$h_A = 13m$
$h_B = 0\ m$

$v_B = \sqrt{2gh_A}$
$v_B = \sqrt{2(10)(13)} = 16.12\ m/s$

$N_B = mg + \dfrac{mv^2}{r^2}$

$= 30\ kg\left(10\dfrac{m}{s^2}\right) + \dfrac{30\ kg\left(16.12\dfrac{m}{s}\right)^2}{15^2}$

$= 300 + 34.65$

$\boxed{= 334.65 N}$

The ride will be safe for Ryan

    We also found that students didn't necessarily think of Newton's 2$^{nd}$ Law as a vector equation. In addition, even if students knew that both the solved and quiz problems were dealing with centripetal acceleration, which is a vector, they didn't necessarily notice the difference between the two (one is at the top of the circle; the other is at the bottom) on their own. When the researcher asked student A to explain how he got the minus sign in his final answer of $T = -\dfrac{mv^2}{r} + mg$ in the tire swing problem, he answered:

*Student A: Isn't that the same as this [pointing to the solved problem] … wait…'cause the centripetal acceleration is going… Wait… No… No, I was wrong. Wait a second. This time I'm on the top, not the bottom… so instead of negative, the centripetal acceleration will be positive, correct?*

When the researcher later asked him to reflect on his learning, he also mentioned that:

*Student A: At first I thought they were just the same situation. I just kind of assumed that they were. I forgot that this one is at the bottom. So I just used whatever I knew from here. It wasn't right.*

    The conversation above suggests that even though the student may have all the physics knowledge required to answer the quiz problem, the knowledge might not be

structured in a well-organized manner to allow him to quickly detect the difference between the two situations [quiz problem and solved problem]. More specifically, it's possible that the connection between Newton's 2$^{nd}$ Law and its vector nature (which implies that the direction of the net force and the acceleration should be contemplated carefully) was not strong enough in the student's mind. Therefore, the student didn't realize that a modification in the application detail should be made in the new situation until additional guidance which directed his attention to this issue was explicitly provided by the researcher.

As pointed out in the section on written quantitative data, some interviewed students were also not consistent while answering different parts of the quiz. The following conservation with student D is an example. Although student D's answer to the dialogue question in intervention 3 was correct and he didn't think there would be an extra force in a circular motion, he later said that he was thinking about what the 1$^{st}$ person in the dialogue question said (which was wrong) when he was asked to explain how he obtained his answer ($T = \dfrac{mv^2}{r}$) for the tire swing problem.

*Student D: [reading the dialogue problem]… I'd agree with person two just because I don't think that… uh… I don't' think it's an extra force. I know that centripetal force is what keeps it going in the circle… but I don't think it's an extra…or is it? Uh… mg… N-mg equal… No. I agree with person two. I don't think… I think… I don't think it's an extra force at point A.*

*Researcher: So… when you wrote down this one [$T = F_c = \dfrac{mv^2}{r}$, his answer to the quiz problem], can you tell me which principle you were using?*

*Student D: Tension is equal to the centripetal force if there's … No I think it's almost wrong… but… I think maybe I was thinking about centripetal force… no I was not thinking about centripetal force at all……*

*[Student D tried to solve the quiz problem again, this time using $F = ma$, $N - W = ma$. (He later noted that what he had as N was in fact the tension.) After he came up with the correct answer for tension, he noted the following]*

*Student D: This [his original work] is wrong. I was just thinking about the centripetal force just because… because of the part A [pointing to what the 1$^{st}$ person in the dialogue question said, which is incorrect.]*

We can invoke the knowledge in pieces [59,60] framework to understand the student's response. The conservation above suggests that Student D had some relevant knowledge but the student's knowledge was not organized in a knowledge structure and he didn't notice the inconsistency between different knowledge elements he referred to unless explicitly guided.

Although the dialogue in intervention 3 didn't necessarily help all students, the interview with student A suggests that the dialogue could be useful for helping students learn the concept of centripetal force if the student tries to incorporate the newly acquired knowledge into his original knowledge structure and is made aware of the conflicts between the knowledge he acquired from the quiz activity and his prior knowledge. In the interview with student A, we found that the notion of associating the centripetal force as an additional force coming from a single physical object was strong. The student could correct his own mistake regarding the incorrect sign for the centripetal acceleration term (after realizing that the direction of the acceleration in the quiz problem was not the same as in the solved problem) and came up with the correct equation by following the procedure in the solved problem (first drawing a correct free body diagram (FBD) and then applying Newton's 2nd Law correctly). However, when he later explained why the tension was maximum at the bottom during the ride, the diagram he drew still suggested that he had a tendency to consider the centripetal force as an additional force coming from a physical object. Fig 3 and Fig 4 show the different diagrams he drew to solve for the tension force and to explain why tension would be maximum at the bottom of the ride, respectively. When he later compared his new figure (Fig 4) to his final answer for tension ($T = mg + \dfrac{mv^2}{r}$), he became confused because in his diagram, $ma_c$ and $mg$ pointed in different directions but in the equation they were added together.

Fig 3 The diagram student A drew from which he came up with the correct answer for tension.

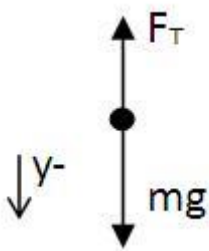

Fig 4 The diagram student A later drew which implied that he was thinking of centripetal acceleration as an additional force.

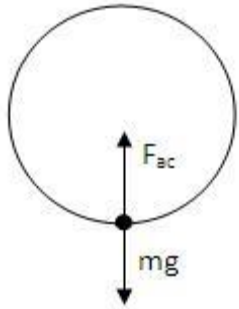

In order to help student A, the researcher discussed with the student the implications of considering the centripetal force as an additional force versus the net force. This discussion was very similar to the information presented in the dialogue question in intervention group 3 except that the case the researcher discussed was for an object at the bottom of the circle instead of at the top of the circle. After the discussion about the two different diagrams and the corresponding equations (similar to those presented for intervention 3 in the supplementary material) the student realized why he had difficulty. He changed his labeling of $F_{ac}$ on Fig 4 to $F_T$, and had the following conversation with the researcher:

Student A: I see what I was doing wrong. I was confused about that. Now it makes more sense.

Researcher: Yeah, but still I don't understand how I helped you. So, can you explain more?

Student A: Yeah, you helped me because I guess I was thinking of this [$F_{ac}$] as a force, like as a physical force, so I put it up this way [pointing to his new diagram of Fig 4]. And then I'm really confused because they are acting in two different directions.

Researcher: Yeah but still you use…

Student A: Yeah. But when I originally did it, I just wrote this [pointing to his original diagram of Fig 3], which makes more sense, because my $F_T$ minus mg equals this. So whenever you make the equation, you end up you're adding them

Researcher: OK

Student A: 'cause there… [sigh…] why… or another word is [that they are] acting in the same direction… I… I just got confused by thinking of the ma part as… not the net force but as the… like force acting on that [the object] like that. So whenever you put $ma_c$ equals that [the net force, $F_{ac}$] and then use

*Newton's 2$^{nd}$ Law, it makes a lot more sense to me.*

*Researcher: Yeah, so I think that's another reason why I prefer to draw the acceleration… I mean, beside, not on the…*

*Student A: Yeah, not like direct on that because it confuses [me]*

The discussion above suggests that the dialogue problem and the related concepts presented in intervention 3 can be used as a tool to help students understand the centripetal force. Moreover, it would also be helpful to explicitly require students to draw the acceleration on the side of a FBD (but not directly with other forces). Overall, based on the interviews, we found that if students were actively engaged in the thinking process and if sufficient scaffolding support was provided to help them contemplate the following issues, they were very likely to solve the quiz problem correctly: (1) They realized that the centripetal force is just a name given to one component of the "net force" in a circular motion. It is not always associated with a single physical force unless only one force is present in that direction. (2) They knew how to find the acceleration and its direction. They also discovered that the positions of the objects (relative to the circles) are different in the two problems since one object is at the top and the other is at the bottom. (3) They realized how to use Newton's 2$^{nd}$ law correctly as a vector equation instead of as a scalar equation. (4) They were required to draw an arrow indicating the direction of the acceleration not on the FBD but on the side of it. Follow-up studies including interviews with students from all levels of expertise could be conducted in the future to thoroughly explore the specific effects different scaffolding supports could have on each of these issues.

## **Summary and Future Outlook**

In this study, we found that students in both the calculus-based and algebra-based courses were able to recognize the similarities between the isomorphic problems in terms of the relevant physics principles involved when they were asked to learn from a solved problem and transfer what they learned from the example problem to solve another isomorphic quiz problem. The algebra-based students in all three intervention groups on average outperformed the comparison group students in the same course because many of them had no clue about how to approach the quiz problem if no support was provided. Providing algebra-based students with a solved isomorphic problem to learn from (regardless of the types of additional scaffolding supports involved in three different intervention groups) improved their performance by helping them invoke the relevant principles in the quiz problem. On the other hand, students in the calculus-based course were better than the algebra-based students in the sense that even without the solved problem provided, they already had some idea about the structure of the problem, although they may not have been able to proceduralize the principles correctly. Therefore, a significant improvement would be observed if the students were not only able to identify the similar principles involved in the two problems, but were also capable of applying what they learned from the solved example in an appropriate way to the new situation presented in the quiz problem. Among all three interventions, we found that intervention 2, in which students were asked to try the quiz problem on their own before the solved problem was provided, was the best intervention in helping the calculus-based students. The findings suggest that postponing the scaffolding support until students have attempted to solve the quiz problem without help is consistently beneficial for students in both courses because the clear targeted goal and the thinking process students went through in their first attempt facilitates better transfer to the other problem.

As noted earlier, the greatest difficulty students had in the analogical reasoning activity discussed was in the correct application of the principles in the new context. One common difficulty observed, for example, was that many students failed to differentiate between the situations in which an object is going over the top versus the bottom of a circle and they didn't contemplate the direction of the corresponding centripetal acceleration and its sign in the corresponding equation. In general, calculus-based students performed better than the algebra-based students on the transfer problems.

In order to help students perform better on the transfer problem, more scaffolding may be required. Deliberately guiding students to think more about the relations between the isomorphic problems by helping them discern not only the similarities, but also the differences between the isomorphic problems and asking them to discuss the implications of both the similarities and differences before actually solving the transfer problem may be a useful strategy. It is possible that by performing a systematic and thorough comparison of the two problems, students may think through the analogy more comprehensively and carefully. If students are new to such activities and they have difficulty identifying the differences they should be looking for in the isomorphic problems, other strategies that are helpful for learning such as instructor modeling, peer discussion, etc. may be combined to assist students (at least in the beginning). It is likely that with more practice and feedback on such analogical reasoning activity, students will gradually develop expertise. The scaffolding support can be reduced as the students develop self-reliance.

A similar strategy to assist students in discerning the differences between the problems and contemplating the application details is to provide them with more than one solved problem to learn from. If two isomorphic solved problems which contain different contexts and different application details are provided to them, students can no longer simply match the quiz problem to either one of them without thinking. They will have to carefully examine the similarities and differences between the three problems and combine what they learned from both solved problems to come up with a new solution that is suitable for the quiz problem. The different application details presented in the two solutions could also serve as a model and/or a hint for how different situations may require the application of the same principles but the application details must be adjusted in each situation.

Some additional scaffolding supports could be designed (and may be combined with the previous strategies) to help students with specific difficulties. For example, one common difficulty found in students' work on the quiz problem was that they didn't draw a free-body diagram when solving the quiz problem. It is possible that mistakes related to missing the gravitational force or having an incorrect sign for the acceleration term (as described in the results section) could be reduced if, in addition to the current intervention, students are explicitly asked to draw a free body diagram before solving the problem, and a comparison between the free body diagrams for the tire swing problem and the roller coaster problem is explicitly enforced. It is also useful to help students develop the habit of drawing the acceleration on the side of the FBD as discussed in the interview. The acceleration vector drawn on the side may help remind students about the fact that they have to consider the vector nature of both forces and accelerations when applying Newton's $2^{nd}$ law. At the same time, it avoids the difficulty of students confusing the centripetal force as an additional force if the arrow signifying the acceleration is drawn together with all the forces.

In summary, deliberately using isomorphic worked out examples to help students transfer what they learned from one context to another can be a useful tool to help students understand the applicability of physics principles in diverse situations and develop a coherent knowledge structure of physics. For introductory students, such well-thought out activity could provide a model for effective physics learning since the idea of looking at deep similarities beyond the surface features is enforced throughout the activity. It is possible that students will become more facile at the analogical reasoning processes if practice and feedback are constantly provided to them. The greatest benefit may be achieved if similar activities are sustained throughout the course over different topics and the coherence of physics as well as the importance of looking at the deep features of the problems is consistently explained, emphasized, demonstrated and rewarded by the instructors.